# What would be outcome of a Big Crunch?

Dragan Slavkov Hajdukovic[1]
PH Division CERN
CH-1211 Geneva 23
dragan.hajdukovic@cern.ch
[1]On leave from Cetinje; Montenegro

**Abstract**
I suggest the existence of a still undiscovered interaction: repulsion between matter and antimatter. The simplest and the most elegant candidate for such a force is gravitational repulsion between particles and antiparticles. I argue that such a force may give birth to a new Universe; by transforming an eventual Big Crunch of our Universe, to an event similar to Big Bang. In fact, when a collapsing Universe is reduced to a supermassive black hole of a small size, a very strong field of the conjectured force may create particle-antiparticle pairs from the surrounding quantum vacuum. The amount of antimatter created from the physical vacuum is equal to the decrease of mass of "black hole Universe" and violently repelled from it. When the size of the black hole is sufficiently small, the creation of antimatter may become so huge and fast, that matter of our Universe may disappear in a fraction of the Planck time. So fast transformation of matter to antimatter may look like a Big Bang with initial size about 30 orders of magnitude greater than the Planck length, questioning the need for inflation. In addition, a Big Crunch, of a Universe dominated by matter, leads to a new Universe dominated by antimatter, and vice versa; without need to invoke CP violation as explanation of matter-antimatter asymmetry. Simply, our present day Universe is dominated by matter, because the previous Universe was dominated by antimatter.

The prevailing opinion in the contemporary Cosmology is that our Universe was born in a Big Bang. One of possibilities is that it will end in a Big Crunch; i.e. gravitational collapse of a gigantesque mass estimated to be of the order of $10^{52} kg$. If so, the major question is what would be outcome of a Big Crunch.

Our actual knowledge is so incomplete, that a satisfactory scientific answer on this question can't be given. However, when the rigorous scientific approach is impossible, it is the best time for imagination and speculations. I like words of Einstein: "imagination is more important than knowledge". Imagination and speculations is often a key, opening the door for a new quantum leap in science.

Let's start by noting, that, in the final stage of a Big Crunch, Universe is reduced to a suppermassive black hole with a linear size, many orders of magnitude smaller than the corresponding Schwarzschild radius.

Now, let's imagine that there is a still undiscovered interaction: repulsion between matter and antimatter, with the following features:

- o  The new force acts, between particles having an appropriate "charge", which (just as the electric charge) can be positive or negative. In order to be definite, a positive "charge" is attributed to matter and a negative one to antimatter.
- o  There is an attractive force between "charges" of the same sign, and a repulsive force between "charges" of different sign. This is just opposite to the familiar case of electric charges.
- o  It may be, but it is not required to be, a long-range force. In fact, when the size of the collapsing Universe is smaller than the range of interaction, a short-range force is de facto a long-range one.
- o  It may be, but it is not required to be, a universal force like gravitation, acting between all particles and antiparticles. It may be a non-universal force, limited to a class of particles (just as the Coulomb force is limited to particles with electric charge).

Let us point out, that the conjectured repulsion between particles and antiparticles may be a new force, but in fact, the most elegant and economical solution is to assume, that it is not a new force, but a generalization of gravity, with included gravitational repulsion between matter and antimatter. Without entering the complex discussions (and the appropriate modifications of General Relativity), the simplest way to define such a gravitational interaction is to assume simultaneous validity of the Newton law of gravity and the following relations:

$$m_i = m_g \,;\ m_i = \overline{m}_i \,;\ m_g + \overline{m}_g = 0 \tag{1}$$

Here, a symbol with a bar denotes antiparticles; while indices $i$ and $g$ refer to inertial and gravitational mass. The first two relations in (1) are experimental evidence [1], while the third one is my conjecture which dramatically differs from general conviction $m_g - \overline{m}_g = 0$.

While I favour the gravitational repulsion between matter and antimatter, let us allow possibility of a new force. In such a case it is assumed that the conjectured repulsion between particles and antiparticles dominates the gravitational attraction between them.

In order to understand the physical significance of the above conjecture, let's start with illuminating example coming from the Quantum Electrodynamics (QED): creation of electron-positron pairs from the (Dirac) vacuum by an external (classical i.e. unquantized), constant and homogenous electric field $E$. In this particular case, the particle creation rate per unit volume and time is known exactly [2], and may be written in the form:

$$\frac{dN_{e^+e^-}}{dtdV} = \frac{4}{\pi^2} \frac{c}{\lambdabar_e^4} \left(\frac{E}{E_{cr}}\right)^2 \sum_{n=1}^{\infty} \frac{1}{n^2} \exp\left(-\frac{n\pi}{2} \frac{E_{cr}}{E}\right) \tag{2}$$

where

$$\lambdabar_e = \frac{\hbar}{m_e c} \quad and \quad E_{cr} = \frac{2 m_e^2 c^3}{e \hbar} \tag{3}$$

are respectively the reduced Compton wavelength of the electron, and the critical electric field. For the purpose of the present paper it is useful to replace quotient of electric fields $E/E_{cr}$ with the quotient of corresponding accelerations $a/a_{cr}$ and to allow particle-antiparticle pairs with any mass $m$. Hence instead of Eq.(2) we have equation:

$$\frac{dN_{m\overline{m}}}{dtdV} = \frac{4}{\pi^2} \frac{c}{\lambdabar_m^4} \left(\frac{a}{a_{cr}}\right)^2 \sum_{n=1}^{\infty} \frac{1}{n^2} \exp\left(-\frac{n\pi}{2} \frac{a_{cr}}{a}\right) \,;\ a_{cr} = \frac{2c^2}{\lambdabar_m} \tag{4}$$

that may be used not only in the case of an electric field, but also in the case of any other field attempting to separate particles and antiparticles. Let's note that acceleration $a \equiv a(m, R)$ is a function of both: mass $m$ and position $R$ of created particles.

As well known, the above phenomenon is due to both, the complex structure of the physical vacuum in QED and the existence of an external field. In the (Dirac) vacuum of QED, short-living "virtual" electron-positron pairs are continuously created and annihilated again by quantum fluctuations. A "virtual" pair can be converted into real electron-positron pair only in the presence of a strong external field, which can spatially separate electrons and positrons, by pushing them in opposite directions, as it does an electric field $E$. Thus, "virtual" pairs are spatially separated and converted into real pairs by the expenditure of the external field energy. For this to become possible, the potential energy has to vary by an amount $eE\Delta l > 2m_e c^2$ in the range of about one Compton wavelength $\Delta l = \hbar/m_e c$, which leads to the conclusion that the pair creation occurs only in a very strong external field, greater than the critical value $E_{cr}$ in Equation (3).

My conjecture is tailored in such a way, that, deep inside the horizon of a black hole, the field of the assumed force between matter and antimatter can create particle-antiparticle pairs from the physical vacuum; with the additional feature that a black hole made from matter violently repels the created antiparticles, while a black hole made from antimatter violently repels particles. Without lost of generality we may limit to the case of a black hole made from matter.

Now, the qualitative picture is very simple. The amount of created (and violently repelled) antimatter is equal to decrease in the mass of black hole. Hence, during a Big Crunch, quantity of matter decreases while quantity of antimatter increases for the same amount. If this process of transformation of matter to antimatter is very fast it may look as a Big Bang.

For simplicity, as a toy model, let's consider black hole as a miniscule ball with radius $R_H$. While $R_H$ decreases, the acceleration $a(m, R_H)$ at the "surface" of black hole increases and in principle particle-antiparticle creation rate (4) may become so huge to transform matter in antimatter in a time interval as small as the Planck time.

The most important part of this qualitative picture is that Big Crunch of a Universe made from matter, leads to a Big Bang like birth of a new Universe made from antimatter. Hence, the question why our Universe is dominated by matter has a simple and striking answer: because the previous Universe was made from antimatter. There is no need to invoke CP violation as explanation of the asymmetry between matter and antimatter.

This beautiful qualitative picture is valid in the case of a gravitational repulsion between matter and antimatter. However it is also valid in the general case of a new repulsive force between matter and antimatter if it dominates the eventual gravitational attraction between them. This assumption allows determining a lower-bound for particle-antiparticle creation rate (4) in the general case. In fact, it is easy to estimate particle-antiparticle creation rate in the particular case $m_g + \overline{m}_g = 0$ and in the same time it may serve as estimation for the lower-bound in all other cases. It is just what is done in the rest of this paper; the following relations (6), (7), (8), (9) and (10) correspond to the case of gravitational repulsion between matter and antimatter but they can also be considered as lower-bound in the more general case.

Firstly, let's define a critical radius $R_{Cm}$ as the distance at which gravitational acceleration has the critical value $a_{cr} = 2c^2/\lambdabar_m$, defined by Eq.(4).. Combining value for $a_{cr}$ with the Newton's law of gravitation leads to

$$R_{Cm} = \frac{1}{2}\sqrt{\lambdabar_m R_S} \equiv L_P \sqrt{\frac{M}{2m}} \tag{6}$$

where $R_S$ is the Schwarzschild radius and $L_P = \sqrt{\hbar G/c^3}$ is the Planck length. Hence a sphere shell with the inner radius $R_H$ and the outer radius $R_{Cm}$ should be a „factory" for creation of particle-antiparticle pairs with mass $m$. Of course, for a stronger force, $R_{Cm}$ must be larger. As we have used the weakest allowed force, the result (6), under the assumption that the range of the interaction is not smaller than $R_{Cm}$, may be considered as a lower-bond for stronger interactions.

Next, the particle-antiparticle creation rate is significant only for an acceleration $a$ greater than the critical acceleration $a_{cr}$. If $a > a_{cr}$, the infinite sum in Eq. (4) has numerical value not too much different from 1. So, as we are interested only in the order of magnitude, instead of Eq.(4), a simple but good approximation is:

$$\frac{dN_{m\overline{m}}}{dt dV} = \frac{4}{\pi^2} \frac{c}{\lambdabar_m^4} \left(\frac{a}{a_{cr}(m)}\right)^2 \tag{7}$$

Now, after integration, Eq. (7) leads to:

$$\frac{dN_{m\overline{m}}}{dt} = \frac{c}{\pi}\left(\frac{R_S}{\lambdabar_m}\right)^2 \frac{R_{Cm} - R_H}{R_H R_{Cm}} = \frac{4c}{\pi}\left(\frac{Mm}{M_P^2}\right)^2 \frac{R_{Cm} - R_H}{R_H R_{Cm}} \tag{8}$$

When $R_{Cm} >> R_H$ it can be further simplified to

$$\frac{dN_{m\overline{m}}}{dt} = \frac{4c}{\pi}\left(\frac{Mm}{M_P^2}\right)^2 \frac{1}{R_H} \tag{9}$$

Now, let's calculate numerical values, for particle-antiparticle pairs having respectively mass of electron, mass of proton and Planck mass. With the mass of the Universe estimated to be of the order of $10^{52} kg$, Eq.(6) gives the corresponding orders of magnitude:

$$R_{Ce} \sim 10^6 m; \quad R_{Cp} \sim 10^5 m \quad R_{CM_P} \sim 10^{-5} m \tag{9}$$

Let's focus on the possibility of creation of particle-antiparticle pairs with mass equal to Planck mass. The first possibility is that the range of the conjectured interaction is not smaller than the critical radius $R_{CM_P}$. If so, creation of so heavy particles is possible already when the size of the collapsing Universe is about $10^{-5} m$, i.e. 30 orders of magnitude greater than the Planck length! With a choice $R_H \sim 10^{-6} m$, Eq.(9) gives the following orders of magnitude for particle-antiparticle creation rates:

$$\frac{dN_{ee^+}}{dt} \sim 10^{90} \ pairs/s \ ; \quad \frac{dN_{p\bar{p}}}{dt} \sim 10^{96} \ pairs/s \ ; \quad \frac{dN_{M_P}}{dt} \sim 10^{134} \ pairs/s \qquad (10)$$

The last of numerical relations (10), tells us, that decrease of matter and increase of antimatter have rate of $10^{126} kg/s$, while the estimated mass of our Universe is „only" about $10^{52} kg$ ! Hence, these rates are so big, that a Big Crunch of our Universe may be transformed to a Big Bang, in a fraction of Planck time! The size of the new born Universe is more than 30 orders of magnitude greater than the Planck length; what may be an alternative to inflation in Cosmology.

Of course we do not know the range of the conjectured repulsion between matter and antimatter. Hence, the second possibility is, that the range of the force is much smaller than the critical radius $R_{CM_P}$. If so, the new Universe would be born with a smaller size; in fact, with a size approximately equal to the range of the repulsive force between matter and antimatter. But, the most important is, that, whatever the range of interaction, a Big Crunch would be always transformed to a Big Bang. Our Universe would not end as a black hole, but would disappear giving birth to a new Universe. We may live in a Universe which is just one in a long series of Universes, alternatively made from matter and antimatter.

It may be appropriate to name the new force "the birth force" and it would be nice, if it is not a new force, but just a gravitational repulsion between matter and antimatter (existing at least at short distances).. In the near future, a new generation of experiments (AEGIS [3] at CERN and AGE [4] at Fermilab) will test if there is a long-range repulsion between matter and antimatter. Eventual short range repulsion may stay a headache for many generations of experimental physicists.

Let's point out that the existence of the long-living black holes is not necessarily an argument against the above scenario. The mass of the Universe is many orders of magnitude greater than the most massive black holes. Presumably the outcome of a gravitational collapse depends on the initial mass of a system; bellow a critical mass, the collapse ends as a long-living black hole (as argued in Ref. [5]), while above a critical mass, it ends as a Big Bang like event.

In fact, my conjecture concerning the gravitational proprieties of antimatter is just one of two complementary approaches, both suggesting that the role of antimatter is poorly understood and highly underestimated in the contemporary cosmology. The complementary postulate may be summarized as; antimatter repels ordinary matter and also is self-repulsive, while matter attracts both particles and antiparticles. Such a postulate makes plausible coexistence of matter and antimatter in our Universe, and opens attractive possibility (see [6] and [7] and references therein) that what we call dark energy is just a consequence of gravitational repulsion caused by huge quantities of antimatter located at intergalactic voids. In my approach, both, matter and antimatter are self-attracting while there is gravitational repulsion between them. The impact of antimatter is not caused by its presence in the Universe but through interaction of the physical (quantum) vacuum with the ordinary matter ([5], [8]). If one of these complementary (and mutually excluding) postulates is correct it would be a quantum leap in our understanding of the Universe.